\documentclass[final,3p,times,twocolumn]{elsarticle}

\usepackage{amssymb}
\usepackage{amsmath}
\usepackage{color}
\usepackage{array,multirow}
\usepackage{url}

\newcommand{\eref}[1]{Eqn.~\ref{#1}}
\newcommand{\fref}[1]{Fig.~\ref{#1}}
\newcommand{\tref}[1]{Tab.~\ref{#1}}

\journal{Chemical Physics}

\begin{document}

\begin{frontmatter}



\title{A quantum Monte Carlo study of systems with effective core potentials and node nonlinearities}


\author[inst1]{Haihan Zhou}
\ead{hzhou23@ncsu.edu}
\author[inst2]{Anthony Scemama}
\author[inst1]{Guangming Wang}
\author[inst1]{Abdulgani Annaberdiyev}
\author[inst1]{Benjamin Kincaid}
\author[inst2]{Michel Caffarel}
\author[inst1]{Lubos Mitas}

\affiliation[inst1]{organization={Department of Physics},
            addressline={North Carolina State University}, 
            city={Raleigh},
            postcode={27695-8202}, 
            state={North Carolina},
            country={USA}}

\affiliation[inst2]{organization={Laboratoire de Chimie et Physique Quantiques (UMR 5626)},
            addressline={Universit\'e de Toulouse, CNRS, UPS}, 
            country={France}}
            


\begin{abstract}
We study beryllium dihydride (BeH$_2$) and acetylene (C$_2$H$_2$) molecules using real-space diffusion Monte Carlo (DMC) method. The molecules serve as perhaps the simplest prototypes that illustrate the
difficulties with biases in the fixed-node DMC calculations that might appear  with the use of effective core potentials (ECPs) or other nonlocal operators.
This is especially relevant for the recently introduced correlation consistent ECPs (ccECPs) for $2s2p$ elements. Corresponding ccECPs exhibit deeper potential functions due to higher fidelity to all-electron counterparts, which could lead to larger local energy fluctuations.
We point out that the difficulties stem from issues that are straightforward to address by upgrades of basis sets,  use of T-moves for nonlocal terms,
inclusion of a few configurations into the trial function and similar. 
The resulting accuracy corresponds to the ccECP target (chemical accuracy) and it is in consistent agreement with independent correlated calculations. Further possibilities for upgrading the reliability of the DMC algorithm 
and considerations for better adapted and more robust Jastrow factors are discussed as well.


\end{abstract}

%


\end{frontmatter}



\section{Introduction}
\label{sec:intro}

Over the past few decades, quantum Monte Carlo (QMC) methods have provided a promising route for solving many-body problems in a variety of quantum systems. 
In particular, real-space fixed-node (FN) diffusion Monte Carlo (DMC) approaches have demonstrated great accuracy in electronic structure calculations of molecular and solid systems \cite{foulkes_quantum_2001, huntQuantumMonteCarlo2018, kolorencApplicationsQuantumMonte2011}.
In numerous cases, even single-reference trial wave function DMC calculations have obtained results on par with best many-body wave function approaches such as the ``gold standard of quantum chemistry" Coupled Cluster with singles, doubles, and perturbative triples (CCSD(T)) \cite{al-hamdaniInteractionsLargeMolecules2021, al-hamdaniWaterBNDoped2014, dubeckyQuantumMonteCarlo2013}.
The obtained observables such as binding energies, excitation gaps, vibrational frequencies have also been accurate and agreed very well with CCSD(T) results as well as with experiment \cite{wang_performance_2019}.
In addition, DMC and related methods are applicable to much larger systems such as periodic models of solids in both insulating and metallic states, 2D materials, and systems with other effective interactions such ultracold atoms and beyond \cite{zhengComputationCorrelatedMetalInsulator2015, huang_bandgaps_2021, wines_first-principles_2020, shinOptimizedStructureElectronic2021, liAtomicFermiGas2011}.

One of the important ingredients of the DMC electronic structure calculations is the speed-up from  elimination of core states that leads  to valence-only calculations. This is commonly
accomplished by introducing effective core potentials (ECPs) or closely related pseudopotentials,
that open opportunities to study systems with hundreds of valence electrons. Very recently, a new generation of correlation consistent ECPs (ccECPs) has been introduced \cite{bennett_new_2017, bennettNewGenerationEffective2018, annaberdiyevNewGenerationEffective2018, wang_new_2019, annaberdiyevAccurateAtomicCorrelation2020} with the emphasis on accuracy and transferability within the many-body framework. Our construction  of ccECPs relies on several key principles such as 
i) reproducing fully correlated atomic excitations both in all-electron and ECP setting; 
ii) capturing the impact of core-core and core-valence correlations on the valence space;
iii) reproducing hydride and oxide molecular binding curves in order to expand the transferability for non-equilibrium geometries. 
Consequently, the fidelity of ccECPs regarding all-electron calculations is significantly better, and in many cases the accuracy is even beyond frozen core (or uncorrelated core) all-electron calculations.

However, the use of ECPs in DMC and related approaches is not completely
straightforward due to 
nonlocality of the atomic $lm$-projection operators.
This contrasts with the local character of the DMC sampling process, which involves also enforcing the fixed-node constraint. 
In order to overcome this complication, several techniques for incorporating ECPs have been proposed
\cite{mitas_nonlocal_1991, casula_beyond_2006, casula_size-consistent_2010, caffarelUsingCIPSINodes2016} as briefly outlined below.
One is based on projection of the nonlocal operators on the appropriately accurate trial function \cite{mitas_nonlocal_1991}. 
This does not guarantee the upper bound, however, it upholds
the zero variance principle and therefore indirectly enables one to restore the effective convergence towards the exact eigenvalue.
Another way to approach it 
is based on T-moves \cite{casula_beyond_2006, casula_size-consistent_2010} which directly restores the upper bound at the price of not guaranteeing the zero variance directly. 
Here the zero variance is restored indirectly assuming the trial function converges 
to an exact eigenstate, thus the two approaches 
appear to be complementary to each other. 
See also a very recent
development that unifies the advantages of both with gains in efficiency \cite{anderson_nonlocal_2021}.

In some cases, the higher accuracy of ccECPs have led to potential functions for particular symmetry channels that are deeper, i.e., closer to the original (screened) Coulomb interaction than in 
some other commonly used ECP sets. 
In particular, for early  $2s2p$ elements the corresponding $2p-$channel potential is more attractive (as it is also in reality) since this properly reflects the absence of $2p$-core states. 
That could increase 
demands on the quality of trial functions
as well on the accuracy of DMC propagation algorithms. Otherwise, 
the DMC evolution could show significant bias 
or even instability that can prevent reliable energy estimation.
Here we are mainly focused on the impact of particular types of inaccuracies (inadequate basis sets,  trial functions with and without Jastrow factors,
treatment of ECPs) that can cause related effects such as ``runaway" energy 
estimators or walker population instabilities. Such undesirable 
effects can occur due to a combination 
of reasons such as presence of regions with large energy fluctuations combined with 
low probability sampling (i.e. ``stuck walkers") or related inaccuracies in walker propagation
\cite{umrigar_diffusion_1993}. 
This is especially of interest, since it might be rather difficult and/or costly to control this with commonly used techniques such as scanning the time step length and subsequent extrapolations.

In this work, we would like to shed light on these issues, identify the situations in which they typically occur, and point out strategies to 
eliminate the corresponding difficulties.
This will be useful to understand for the cases when the common route of DMC with single-reference trial functions should be used with proper care and whenever large bias or instability is encountered
one has to improve the accuracy of the DMC propagation, improve the trial function, change the treatment of ECPs or even all of the above. 
In particular, we point out examples such as BeH$_2$ and C$_2$H$_2$ 
molecules where we identified such instabilities and demonstrate the ways of getting them under the control. 

This paper is structured as follows.
In section \ref{sec:biases}, we outline the biases regarding the localization of nonlocal ECPs.
Section \ref{sec:methods} discusses the details of the employed methods.
In section \ref{sec:results}, we present the data and elaborate on the results.
Finally, sections \ref{sec:discussion} and \ref{sec:conclusions} includes discussions and concluding remarks.

\section{Nonlocal ECP terms} 
\label{sec:biases}





In the FNDMC method, the node (zero locus) of the trial function is used as the boundary for solving the Schr\"odinger equation in imaginary time.
As it is well-known, the fixed-node condition enables us to avoid the inefficiency of dealing with the fermion sign problem. Denoting the product of the trial function and the FN solution product as $f=\Psi_T({\mathbf R})\Phi_{FN}({\mathbf R},\tau)$, it obeys the following equation
\begin{equation}
    f(\mathbf{R},\tau +\delta \tau) = \int \textrm{d}\mathbf{R}'\; G(\mathbf{R},\mathbf{R}';\delta\tau) f(\mathbf{R}',\tau).
    \label{eqn:f_integral}
\end{equation}
%
The Green's function is approximated for small time steps $\delta \tau$ is given by
\begin{eqnarray}
    G(\mathbf{R},\mathbf{R}';\delta \tau) = (2\pi\delta\tau)^{-3N/2}\times\nonumber\\
   \times \exp\left[  \frac{-\left|\mathbf{R}-\mathbf{R}'-\delta\tau \nabla_\mathbf{R'}\ln \Psi_T \right|^2}{2\delta\tau} \right]\times \nonumber\\
    \times \exp\left[ -\frac{\delta \tau}{2}\left(E_L(\mathbf{R})+E_L(\mathbf{R}') - 2E_T\right) \right] +{\cal O}[(\delta\tau)^3]
    \label{eqn:greens_fn}
\end{eqnarray}
where $N$ denotes the number of electrons and
$E_T$ is the energy offset that allows the number of walkers (normalization of the wave function) to remain asymptotically constant. The local energy is given by $E_L(\mathbf{R})=[H\Psi_T]/\Psi_T$ for the local terms in the Hamiltonian. Its variational average and variance are given by 
\begin{equation}
E_{VMC}=\langle E \rangle=\langle E_L \rangle_{\Psi^2_T}
\end{equation}
\begin{equation}
\sigma^2=\langle (E_L-
\langle E \rangle)^2 \rangle_{\Psi^2_T}
\end{equation}
with analogous expressions using the mixed distribution $f$
instead of $\Psi_T^2$.

Employing ECPs in the DMC approach requires a nontrivial modification of the DMC approach.  
The key point is the presence of nonlocal terms in the corresponding operators and the two most common possibilities how to deal with these parts of the Hamiltonian are outlined below. 


\subsection{Localization approximation}

Atomic ECPs usually involve projectors 
\begin{equation}
W=\sum_{\ell,m}v_{\ell}(r)|\ell m\rangle\langle
\ell m|
\end{equation}
which lead to
off-diagonal (nonlocal) terms
\begin{equation}
W({\bf R},{\bf R}')=\langle {\bf R}|W|{\bf R}'\rangle .
\end{equation}
A propagator with such terms would involve nonlocal jumps that could violate the fixed-node condition. One can see this by expanding
 the Green's function in the limit $\delta\tau\to 0$ 
\begin{equation}
G({\bf R},{\bf R}',\delta\tau)= \delta({\bf R}-{\bf R}')-\delta\tau W({\bf R},{\bf R}') \Psi_T({\bf  R})/\Psi_T({\bf R'}) + ...
\label{G_short}
\end{equation}
It is clear that, depending on the signs of $\Psi_T({\bf R}), \Psi({\bf R}'),$
and $W({\bf R},{\bf R}')$, the Green's function could generate negative probablities
for any finite $\delta\tau$.
In order to avoid this, the localization approximation has been proposed \cite{mitas_nonlocal_1991}.
The nonlocal part of the Hamiltonian is 
evaluated as a projection onto the trial
function that gives an average, many-body effective potential. 
Its contribution to the local energy is given by
\begin{equation}
E_{L}^W({\bf R})=\Psi_T^{-1}({\mathbf R})
\int W({\mathbf R }', {\mathbf R} )\Psi_T(
{\mathbf R}') d{\mathbf R}'
\label{eloc}
\end{equation}
where $W({\mathbf R }, {\mathbf R'} )$ represents the ECPs'
matrix elements from the relevant atom(s).
The resulting local operator 
can cause an energy bias which is quadratic in the difference between
the trial and the exact wave functions (see some recent studies of related bias \cite{dzubak,krogelkent}). 
For sufficiently accurate trial functions, the resulting bias 
is typically smaller or at most comparable to the fixed-node bias.
In addition, the corresponding total energy estimator exhibits zero variance, i.e., in the limit of an exact trial function, the fluctuations vanish as
the second order in the trial function deviation from the exact one. Consequently, for the exact trial function, $E^W_L$ becomes an exact effective potential as well. 
On the other hand, for an approximate trial function the total energy is not necessarily variational due to the presence of the nonexact projection term. 
 
In the Results section (\ref{sec:results}) we will make the distinction between types of 
trial functions used in the projection above,
in particular, whether it includes the Jastrow
factor or not. While most common 
approach is to include the Jastrow, an alternative without Jastrow is a viable route to accurate results as well. In particular, with no Jastrow and Gaussian basis sets, the nonlocal integrals can be carried out explicitly saving thus the computational time while generating a different effective potential and correspondingly modified convergence to the exact trial function limit.
It depends on application which approach might 
be more appropriate for a given problem as explained elsewhere \cite{ginerUsingPerturbativelySelected2013, caffarelUsingCIPSINodes2016}.
We point out this option here for the sake of understanding its impact on studied effects
as presented in the Results section.

\subsection{T-moves} 

An important improvement of the localization approximation was proposed in the form of the so-called T-moves
\cite{casula_beyond_2006, casula_size-consistent_2010}.
In the T-moves approach the action of the nonlocal part of the Green's function is decomposed into
two parts depending on the sign of the importance-sampled operator defined as
\begin{equation}
V({\mathbf R }, {\mathbf R'} )= \frac{\Psi_T({\mathbf R})} {\Psi_T({\mathbf R'})} W({\mathbf R }, {\mathbf R'}),
\end{equation}
which we write as
\begin{equation}
V({\mathbf R }, {\mathbf R'} )= V_+({\mathbf R }, {\mathbf R'} ) - V_-({\mathbf R }, {\mathbf R'}).
\end{equation}
where
\begin{equation}
V_{\pm}({\mathbf R }, {\mathbf R'} )= \frac{ V({\mathbf R }, {\mathbf R'} ) \pm |V({\mathbf R }, {\mathbf R'})|}{2}.
\end{equation}
When ${\mathbf R } \ne {\mathbf R'}$ and $V$ is negative, the nonlocal part of the Green's function is positive (see, \eref{G_short}) and, thus, can be interpreted as a transition probability
density and is sampled
accordingly. In the opposite case, $V=V_+$, the contribution can no longer be sampled. It is then added to the local potential and considered as a weight (branching).
T-moves guarantees the upper bound \cite{ casula_beyond_2006, casula_size-consistent_2010,
tenhaafProofUpperBound1995}; however, due to the sampling of off-diagonal terms, it does not possess the zero variance property even in the case of the exact trial state.
In practice, this involves some decrease in the efficiency; however, overall, the T-moves algorithm is very effective and its use even for large systems is straightforward.
For more details the interested reader is referred to the original works \cite{casula_beyond_2006, casula_size-consistent_2010}.

\section{Computational Methods}
\label{sec:methods}

In order to illustrate the focal points explained above, we choose to study BeH$_2$ and CH$_2$ molecules.
We employed ECPs for all elements and calculations presented in this work.
For hydrogen we used a local, regularized Coulomb ECP denoted as ccECP-reg
that removes $-1/r$ singularity at the origin \cite{wang_new_2019}. Since the radius of the regularized part is rather small, the resulting behavior of the ccECP-reg is very close to the original Coulomb with biases in energy differences 
of the order of a few meV (the bare $1s$
eigenvalue is accurate to $\mu$eV level). 
For Be atom, we probed two different ECPs. 
The first one is semi-local  with He-core (2 valence electrons), which we denote as ccECP/[He] \cite{wang_new_2019}.
The alternative one is again an all-electron, local, regularized ECP (ccECP-reg) where Coulomb singularity at the origin is smoothed out to a finite value
\cite{wang_new_2019}.
For carbon we used a semi-local ccECP/[He] \cite{bennett_new_2017}.
The accuracy of these ECPs has been thoroughly tested as presented 
in the original papers. 
The all-electron ccECP-reg is highly accurate and reproduces not only the valence but also some core properties in a manner that is essentially equivalent to the 
original all-electron Coulomb Hamiltonian.

{\em Trial functions with the Jastrow factor.} We used single- or multi-reference 
Slater-Jastrow trial wave functions
\begin{equation}
\Psi_T({\mathbf R})=\exp(J)\sum_nc_n {\rm det}_n^{\uparrow}[\phi_k({\mathbf r}_i)]
{\rm det}_n^{\downarrow}[\phi_l({\mathbf r}_j)]
\label{eqn:psit}
\end{equation}
where $n$ labels the determinants in the expansion. 
In general, the Jastrow factor $J$ combines one-body electron-ion, two-body electron-electron, and three-body electron-electron-ion  terms and the corresponding expansion coefficients were optimized variationally.  Overall, $\approx$ 140 variational parameters were used in the Jastrow factor expansion.
We used \textsc{Qmcpack} \cite{kim_qmcpack_2018, kent_qmcpack_2020, krogel_nexus_2016} and \textsc{Qwalk} \cite{wagner_qwalk_2009} codes for these QMC calculations.

{\em Trial functions without the Jastrow factor.} In order to probe for the impact of the Jastrow factor on the discussed issues, we also employed 
trial functions without the Jastrow factor, so that $J=0$ in \eref{eqn:psit} as specified in particular calculations below.
This alternative is motivated by important considerations as discussed in \cite{ginerUsingPerturbativelySelected2013, caffarelUsingCIPSINodes2016}.
While the Jastrow factor clearly improves the trial function on average, it might introduce unwanted fluctuations in other areas, typically close to the nucleus, where even a minor modification
of the wave function can unbalance the kinetic and potential contributions. This issue is indeed 
very pertinent for deeper potential functions of ccECPs.
For these types of calculations we used 
\textsc{Quantum Package} \cite{garnironQuantumPackageOpenSource2019} and \textsc{Qmc=Chem} \cite{scemama_qmc} codes.

For independent cross-check of total energies, we used CCSD(T) as well as full Configuration Interaction (FCI) method with complete basis set (CBS) extrapolations. 
The orbitals of the trial functions were expanded in aug-cc-pVTZ basis sets that were designed for our ccECPs \cite{bennett_new_2017, wang_new_2019} using \textsc{Gamess} package \cite{schmidt_general_1993}.
This basis set was enough to reach essentially the CBS limit for the trial function nodes.
Further details can be found in previous papers \cite{foulkes_quantum_2001, kolorencApplicationsQuantumMonte2011} and in Supplementary Material.


\section{Results}
\label{sec:results}

\subsection{BeH$_2$}
\begin{figure}[!htbp] 
    \centering
    \includegraphics[width = 1.0 \columnwidth]{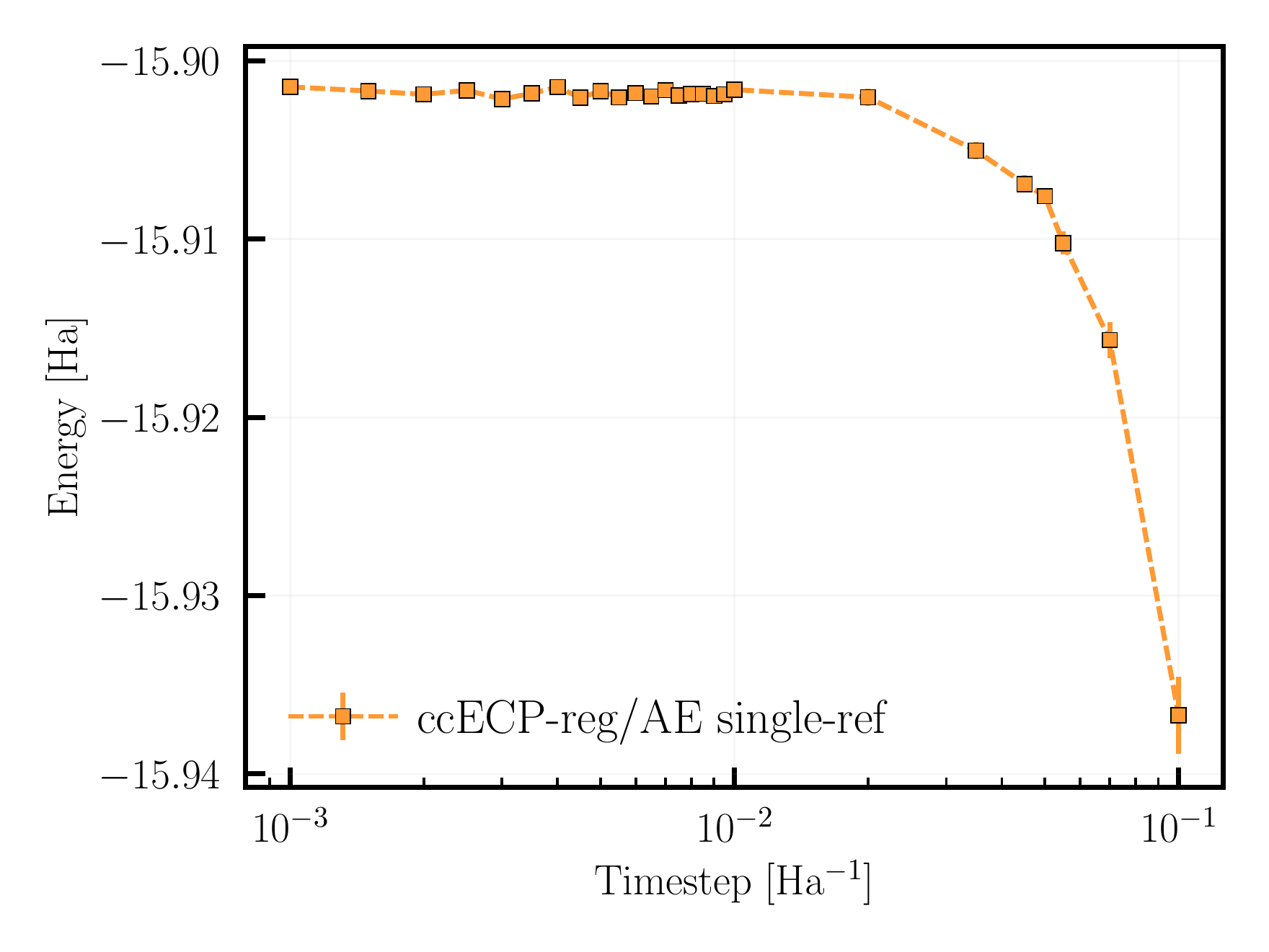}
    \caption{FNDMC energies vs time
    step for BeH$_2$ molecule.
    Be atom is represented by all-electron ccECP-reg/AE. 
    Single-reference trial wave function is used.
Except for the two largest time-steps the statistical fluctuations are negligible and the error bars are smaller than the solid symbols.}
    \label{Fig ae}
\end{figure}

\begin{figure}[!htbp] 
    \centering
    \includegraphics[width = 1.0 \columnwidth]{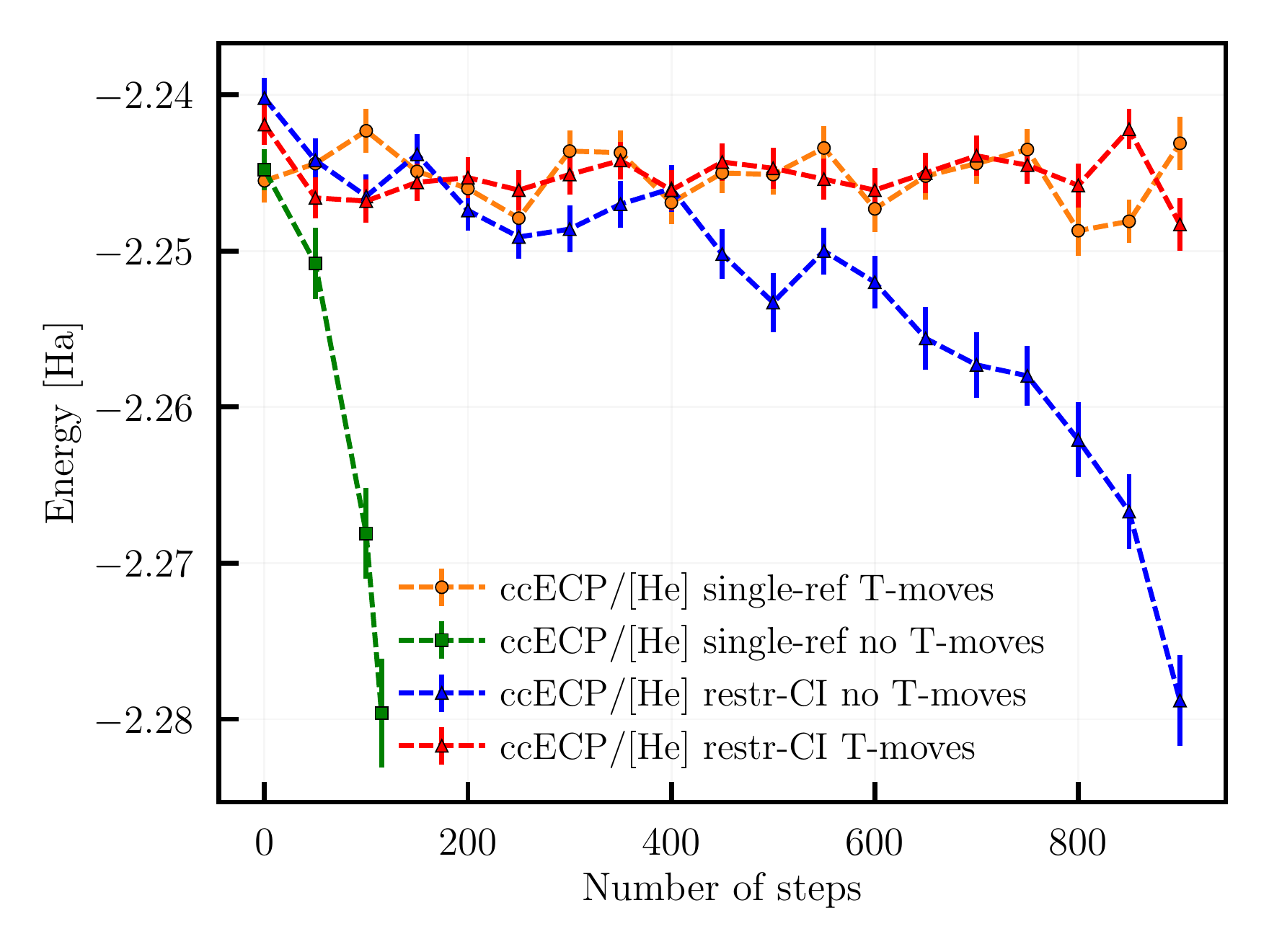}
    \caption{FNDMC energies vs number of steps (in time steps of 0.001 Ha$^{-1}$) for BeH$_2$ molecule.
    Be atom is represented by ccECP/[He].
    }
    \label{Fig qmcall}
\end{figure}

First, we want to recall the well-known fact that the all-electron Be atom with single-reference trial function
shows large fixed-node error \cite{umrigar_diffusion_1993, sharma_spectroscopic_2014, kalemos_nature_2016}
and this persists for ccECP-reg/AE calculation as 
well as for ccECP/[He].
However, the issue  of  
 $2s\to 2p$ degeneracy is alleviated in the BeH$_2$ molecule due to the molecular orbital hybridization that significantly diminishes this effect. 
 Our calculation
with ccECP-reg/AE shows a very typical behavior of the FNDMC energy as a function of time step
with robust behavior at smaller time steps and straightforward extrapolation that eliminates the time step error, see \fref{Fig ae}.
The calculations with ccECP/[He] 
were carried out with two types of trial functions, namely single-reference 
and restricted-CI expansion, which included 
low-lying excitations related
to $2s\to 2p$ effects (altogether 145 configuration state functions (CSFs)).
Surprisingly, these calculations 
show a very different behavior
with  significant biases occuring before one can estimate the 
overall energy average, see \fref{Fig qmcall}.  
In particular, without using T-moves,  both single-
and multi-reference trial functions lead to unstable runs.

This is clearly pointing  out 
that the key issue comes from the presence of the nonlocal terms in ccECP/[He].
Note that in the localization  approximation contribution, \eref{eloc}, one divides by the trial wavefuction 
and that may introduce some strong fluctuations when the trial wavefunction is small.
Note that a  similar situation occurs for the ordinary local energy given by $\frac{H \Psi_T}{\Psi_T}$.
However, in this case
the problem is essentially avoided by using sufficiently good trial wave functions since both numerator and denominator have similar overall
amplitudes at the given sampling point.  
For the localization approximation, 
the situation is not that simple since the numerator is given by the integral, not by the value at the sampling point.  However, in practical calculations, the problematic events related to too low local energies are found to be rare 
and can be handled without significant biases, for example, by using a threshold for large weights
\cite{anderson_nonlocal_2021}.


\begin{figure}[!htbp] 
    \centering
    \includegraphics[width = 1.0 \columnwidth]{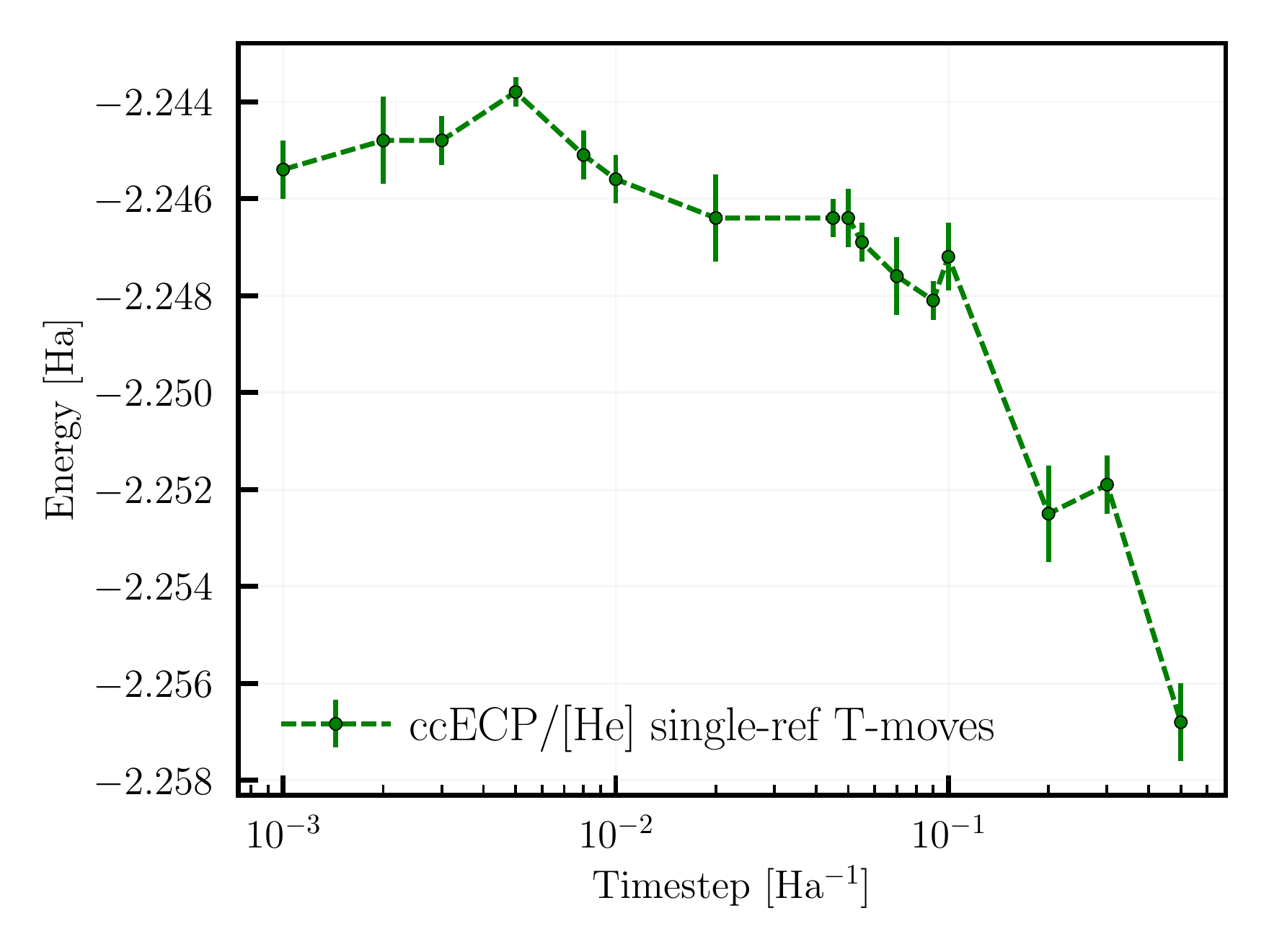}
    \caption{FNDMC energies vs time step for BeH$_2$ molecule. 
    Be atom is represented by ccECP/[He].
    Single-reference trial wave function and T-moves were used. 
    }
    \label{Fig 2e1}
\end{figure}

\begin{table}
\centering
\caption{
Total energies of BeH$_2$ molecule using ccECP/[He] for Be atom.
DMC energies are extrapolated to zero time step and trial functions include the Jastrow factor.
Single-reference and a few-reference (restr-CI/145 CSFs) trial functions that capture the lowest $2s\to 2p$ excitations are shown. 
aug-cc-pVTZ basis set was used to generate the DMC trial wave functions.
FCI/CBS is estimated using a combination of CCSD(T)/CBS and FCI/DZ. See Supplementary Material for details.
}
\label{tab:beh2_totals}
\begin{tabular}{lccl}
\hline
Method      &   $\Psi_T$         & T-moves & Energy [Ha] \\ 
\hline
RHF/CBS     &             &         & -2.16511(1) \\
restr-CI/TZ &             &         & -2.20395    \\
\hline
DMC         & single-ref. & no      & unstable    \\
DMC         & single-ref. & yes     & -2.2452(2)  \\
DMC         & restr-CI    & no      & unstable    \\
DMC         & restr-CI    & yes     & -2.2450(4)  \\
\hline
FCI/CBS     &             &         & -2.2464(1)   \\
\hline
\end{tabular}
\end{table}

The instability is eliminated by using the T-moves as shown in \fref{Fig qmcall} and \fref{Fig 2e1} which illustrate the usual dependence of energy expectations on the time step
and allows for consistent extrapolation.
These obtained energies are summarized in \tref{tab:beh2_totals}.
Note an interesting fact that the single-reference trial function has nodes that are accurate enough so
that even the inclusion of more CSFs such as $2s\to 2p$ configurations leads to the same energy. 
In addition, the resulting values are also very close to the calculated nearly exact results from CCSD(T)/FCI CBS extrapolation. 

Further, we wanted to understand the possible role of the Jastrow factor in this effect. Therefore we have carried out calculations using trial functions without Jastrow so that the projection involves only the antisymmetric Slater component. 
We want to point out that in calculations without the Jastrow factor the localization potential is calculated exactly
by integrals involving only gaussians; therefore, the T-moves option is irrelevant. 
The results turned out to be quite revealing. In particular, the basis set started to play a crucial role and for the DZ basis the calculation became unstable, see \tref{tab:beh2_single_ref}. However, already for TZ basis the energy has stabilized at the consistent value. This clearly points towards an importance of accuracy at one-particle level for correct description of higher values of local kinetic and potential energies that become problematic in ECPs regions. As we can see, rather straightforward adjustments provide straightforward solutions to get this issue under the control.
A related issue was observed before for finite basis sets where spurious nodal artifacts were observed \cite{hachmannNodesHartreeFock2004, bandeRydbergStatesQuantum2006} for Be and C atoms.





\begin{table}
\centering
\caption{
Single-reference DMC energies of the BeH$_2$ molecule using ccECP/[He] for Be and ccECP for H. The trial wave function is a Hartree-Fock determinant, and no Jastrow factor is used. With such a wave function, the integrals involved in the non-local terms have been computed exactly and the T-moves option is irrelevant.
}
\label{tab:beh2_single_ref}
\begin{tabular}{ll}
\hline
Basis & Energy [Ha] \\
\hline
 cc-pVDZ     &    unstable \\
 cc-pVTZ     & -2.24475(6) \\
 cc-pVQZ     & -2.24486(7) \\
 aug-cc-pVDZ & -2.24519(6) \\
 aug-cc-pVTZ & -2.24502(7) \\
 aug-cc-pVQZ & -2.24475(9) \\
\hline
\end{tabular}
\end{table}


\subsection{C$_2$H$_2$}
The ethylene molecule is linear and it has a closed-shell electronic structure with doubly occupied $\sigma(p_z),\pi_x,\pi_y$ molecular orbitals. 
Therefore, all three atomic  $p_x,p_y,p_z$ orbitals are hybridized and the atomic near-degeneracy $2s\to 2p$ effect
in C atoms is therefore basically absent. 
The molecule has a large gap and its triple C$-$C bond is qualitatively similar to the triple bond of nitrogen dimer so that the ground state has a single-reference character and
the trial function should be reasonably well represented by a single configuration. 
Despite this seemingly very straightforward setting, the calculations can show instability similar to the case of BeH$_2$,
see \tref{tab:c2h2_totals}.

\begin{table}
\centering
\caption{
Total energies of C$_2$H$_2$ molecule using ccECP/[He] for C atom.
DMC energies are extrapolated to zero time step and trial functions include the Jastrow factor.
Single-reference and a few-reference (restr-CI/546 CSFs) trial functions that capture the lowest $2s\to 2p$ excitations are shown. 
aug-cc-pVTZ basis set was used to generate the DMC trial wave functions.
FCI/CBS is estimated using a combination of CCSD(T)/CBS binding energy and atomic FCI/CBS \cite{annaberdiyevAccurateAtomicCorrelation2020}.
See Supplementary Material for details.
}
\label{tab:c2h2_totals}
\begin{tabular}{lccl}
\hline
Method      &   $\Psi_T$        & T-moves & Energy [Ha] \\ 
\hline
RHF/CBS     &             &         & -12.1162(1) \\
restr-CI/TZ &             &         & -12.2867    \\
\hline
DMC         & single-ref. & no      & unstable    \\
DMC         & single-ref. & yes     & -12.4632(4) \\
DMC         & restr-CI    & no      & -12.4706(5) \\
DMC         & restr-CI    & yes     & -12.4686(4) \\
\hline
FCI/CBS     &             &         & -12.4817(11) \\
\hline
\end{tabular}
\end{table}

\begin{figure}[!htbp] 
    \centering
    \includegraphics[width = 1.0 \columnwidth]{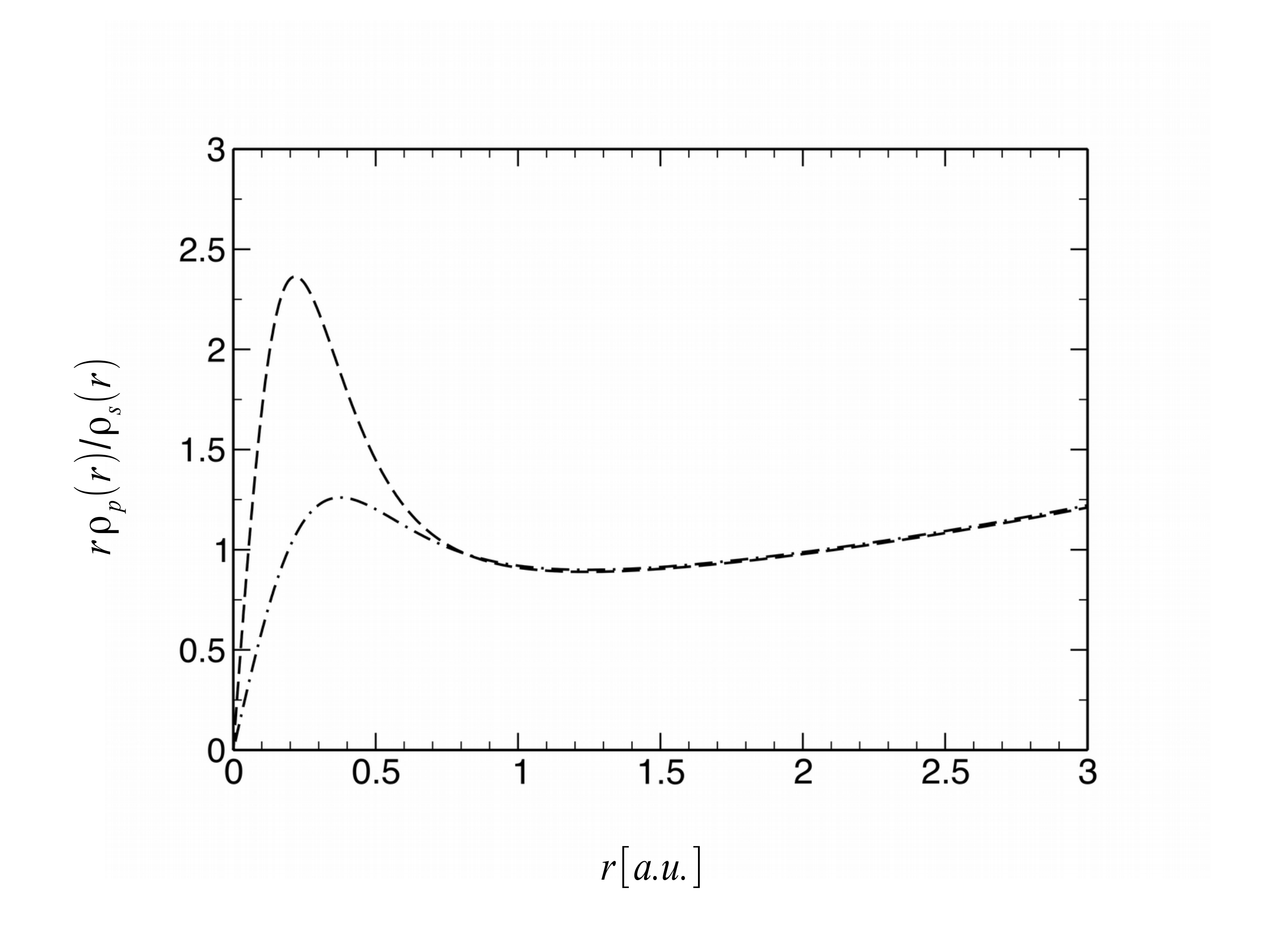}
    \caption{  Ratio of radial atomic (pseudo) orbitals for the C atom from Hartree-Fock calculations. The dashed line corresponds to ccECP while the dashed-dot line is for BFD
    pseudopotential \cite{burkatzkiEnergyconsistentPseudopotentialsQuantum2007}.
    The nonlinearity leads to cavities in the nodal surface (as opposed to linear dependence that would result in a flat node). This feature is present for all $2s2p$ elements, however, it is more pronounced for deeper potential functions and typically for higher ECP accuracy. See the text and Ref. \cite{raschCommunicationFixednodeErrors2014} for further details.}
    \label{Fig catom}
\end{figure}

What is at the root of such complication in this molecular system? 
The reason is the amplitude of the node nonlinearity as it was already 
illustrated previously \cite{raschCommunicationFixednodeErrors2014}.
In the region where the $p-$pseudo-orbital exhibits larger value than the $s-$pseudo-orbital,
the nodal surface becomes strongly curved and shows a
cavity-like shape.
We illustrate schematically this feature in \fref{Fig catom} as a ratio of atomic radial components of $p-$ and $s-$orbitals. 
This is an inherent property of ECP
pseudo-orbitals and it is present in any such ECP resulting in multiple encounters of the node in straight sampling paths as explained previously 
\cite{raschCommunicationFixednodeErrors2014}. In all-electron 
calculations this would occur as well in core hole states, eg, for deep-core ionization or in  excitations from $1s$ to valence
levels. The plotted ratio then shows a diverging maximum due to the radial node of the all-electron $2s$ state. Clearly, 
the nodeless $2s$-pseudo-orbital
smoothes out such divergences to
finite values.

This nonlinearity causes two related effects. 
The first one concerns with the DMC sampling of this region. 
The diffusion-drift propagator includes only gradient to direct the evolution and therefore there is a tendency to generate moves 
that can be often rejected by the Metropolis step 
\cite{reynoldsFixedNodeQuantum1982a, umrigar_diffusion_1993}.
The result is a slow-down of the walker evolution from rejected moves and a decrease of overall efficiency
similar to gradient methods in
banana shaped valleys in multivariate optimizations. 
This could lead to oversampling since the propagation in such regions becomes very different from the rest of the walker ensemble. 
The second complication is similarly problematic,  namely,
the local energy might be strongly deviating from the actual expectation value. 
 Typically, there is a lack of variational freedom that would be flexible enough to properly adjust the wave function and the correlation in such spatially ``tight" regions. 
Note that Jastrow functions are usually rather smooth and they span distances that are larger or at least, comparable to a few bond lengths or several diameters of the atom.
(Also, in solids, the Jastrow factors could involve terms that are long-ranged with perhaps even less short range flexibility.)  
Therefore, wave function nonlinearities with small spatial ranges are very difficult to describe accurately. Improvements through multi-reference expansions might vary in effectiveness and very high excitations might be needed to capture the proper adjustment of the wave function in such regions. 
Thus, if the local energy happens to be very low, the proliferation/weight term could cause difficulties in keeping the population of walkers stable and expectations being free from bias.
If this is further combined with slower local evolution
the resulting process could be unstable.
In general, in order to overcome such biases by
brute force, the sampling might require orders of magnitude smaller time step, which might lead to unreasonable loss of efficiency.

\begin{figure}[!htbp] 
    \centering
    \includegraphics[width = 1.0 \columnwidth]{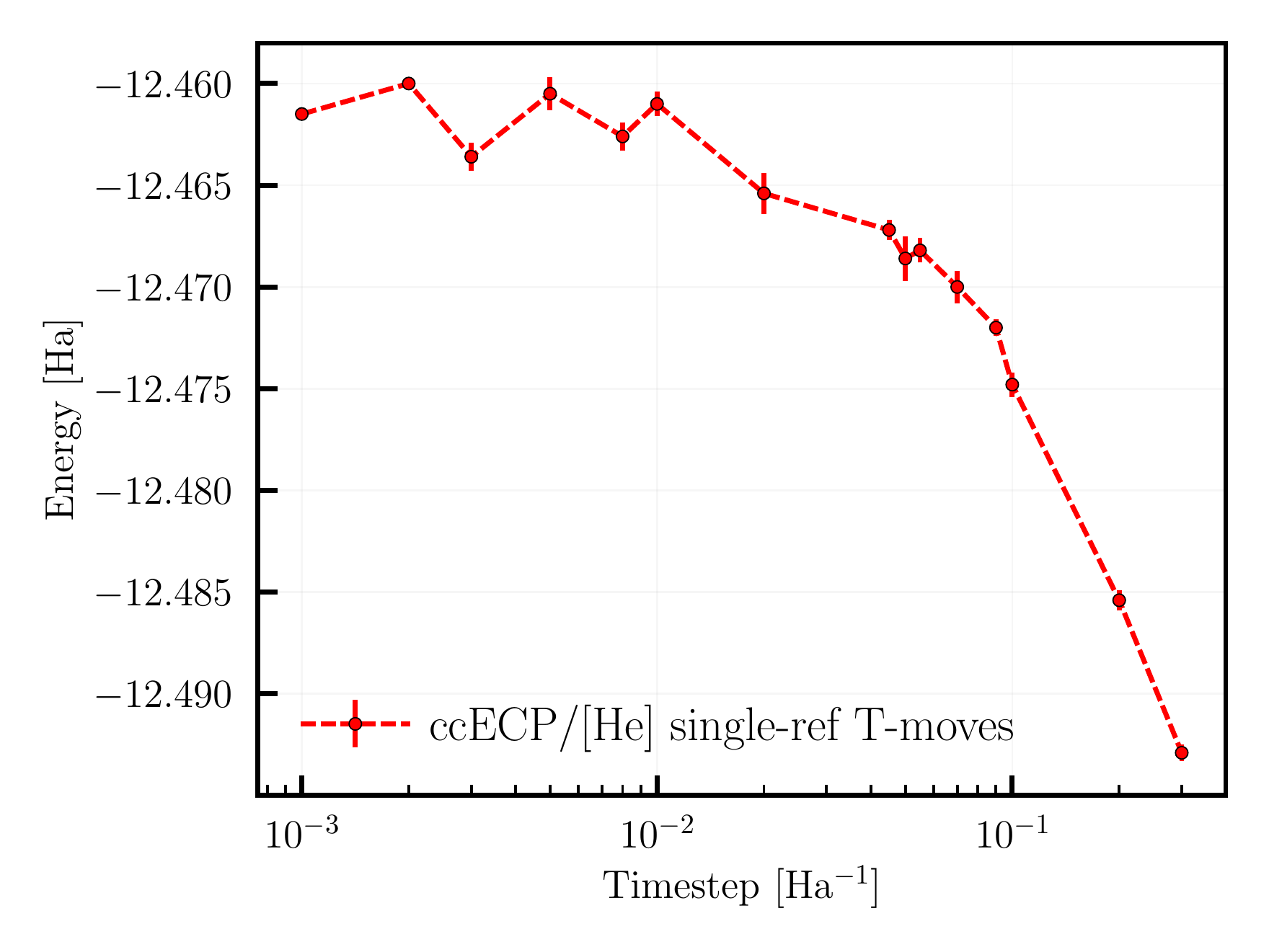}
    \caption{FNDMC energy vs timesteps for C$_2$H$_2$ molecule.
    The C atom is represented by ccECP/[He].
    Single-reference trial wave function and T-moves were used. 
    }
    \label{Fig 2ec2h2}
\end{figure}

As in the previous case, using the option of  T-moves eliminates this problem even for single-reference trial wave functions with the Jastrow factor included, see \tref{tab:c2h2_totals} and \fref{Fig 2ec2h2}. In addition, another obvious possibility is 
taking a multi-reference (546 CSFs) trial function, which also eliminates the bias and 
makes the expectation value of the energy straightforward to calculate in a routine
manner with a very minor 2 mHa difference compared to the T-moves result. This is understandable since without T-moves there is no guarantee that the energy is an upper bound, although we expect any such bias to be smaller than the fixed-node bias.
Comparison with the nearly exact result shows 
that all DMC values without T-moves are an upper bound by significant margin effectively confirming that the leading bias is from the fixed-node condition even with multi-reference trail function (\tref{tab:c2h2_totals}). We expect that
the this would be true uniformly for 
well-optimized trial functions with increasing size of the multi-reference space, i.e., that
the fixed-node bias would dominate all the way
to the exact trial function.

\section{Discussion} 
\label{sec:discussion}

Let us briefly reflect on the obtained results.

{\em Basis sets.}  The role of basis sets in QMC is important for proper description of occupied
and low-lying virtual orbitals which enables one to build the correct nodal structure and local energy landscape without regions with excessive fluctuations.
In particular, 
it is crucial to describe 
areas close to the nucleus with deep potentials and correspondingly large local kinetic energies
with high accuracy. 
In addition, the results show that areas which show higher node curvatures are also  important to be described with sufficient accuracy. Note that due to the wave function continuity its behaviour at the node is one-to-one connected to the exact eigenvalue, see, 
for example, our recent work \cite{mitasWeightedNodalDomain2021}. 
 Once this is achieved and the crucial balance between kinetic and potential 
terms is maintained, the 
sensitivity to basis set becomes less pronounced.
Very similar behavior has been observed also in a very different class of systems such as
noncovalently bonded materials. For example, in van der Waals complexes
the augmentation functions
play a crucial role in  describing the the weak bonding. Once a few augmentation basis functions are included,
the saturation point for QMC methods is reached at intermediate basis set sizes \cite{dubeckyNoncovalentInteractionsQuantum2016, dubeckyQuantumMonteCarlo2013}.

{\em Jastrow factor.} Our results demonstrate perhaps
less appreciated impact of the Jastrow factor on the projection of the nonlocal terms.  While the Jastrow factor is almost universally very effective overall,
 it can cause unwanted distortions in regions with large values of the wave function derivatives that could generate large fluctuations in the local energy.  Besides the solutions we have shown so far, this can be approached in several ways. One possibility is to expand  the Jastrow  variational freedom.
Alternatively,  restricting its spatial reach so as to avoid distorting the wave function in regions with nonlocalities and deep potentials offers another option. 
In fact, some of these issues have been
encountered also before. In preliminary
calculations of systems containing elements with very large cores such as Pt or Au, implementing smooth cut-offs of the Jastrow factors in the core region enabled us to eliminate excessive local energy fluctuations \cite{pt_au}.
This is clearly an area where implementing more flexible Jastrow forms and corresponding benchmark studies would be highly desirable.  

{\em T-moves and DMC propagators.}  
T-moves improve the calculations with nonlocal operators in multiple directions and one of them is the increased robustness \cite{casula_beyond_2006}. 
This is very important from the point of view that this is true even for the simplest one-reference trial functions once the basis set has been saturated. The key observation is that
the large energy fluctuations that affect the 
stability are located
in the same sign domain. Note that
the T-moves impact comes from sampling the nonlocal term for paths that do not cross the node.
Consequently, this suggests that further improvements of the DMC propagator related to the presence of
the nonlocal operator
might be possible and recent work suggests 
interesting options in this direction \cite{anderson_nonlocal_2021}. 
 The DMC method involves the propagation of walkers in position space together with reweighting represented by the local energy factor 
in \eref{eqn:greens_fn}. 
This Green's function should be accurate to the order $(\delta\tau)^3 $ provided the drift and the local energy vary mildly and there are no singularities.
In reality, this is almost never true since the drift diverges at the nodes and related divergences are present in the local energy as well.
These complications have been recognized some time ago and the algorithm was adjusted by
a number of additional improvements. 
This includes, for exmple, the Metropolis rejection step \cite{reynoldsFixedNodeQuantum1982a, umrigar_diffusion_1993} and a smooth cut-off of the drift to a finite value beyond certain threshold \cite{anderson_nonlocal_2021}.
These adjustments modify the walker evolution and effectively improve the corresponding Green's function
in regions that show either very small wave function amplitudes and/or very large fluctuations of local energy. 
The adjustments can uphold not only the upper bound but in fact they provide much more accurate estimations of energy and other observables than
using the plain approximate Green's function from \eref{eqn:greens_fn}. Work in this direction is very promising \cite{anderson_nonlocal_2021}.

\section{Conclusions}
\label{sec:conclusions}
The presented calculations provide a fresh look on the issues that occur in some DMC calculations with ECPs that create regions with large values of kinetic and potential energies and could lead to
significant local energy fluctuations.
This covers some of $2s2p$ elements especially those with small number of valence electrons such as Be, B, and C.
This could occur also for heavier atoms, especially if the density of valence states
in higher angular momentum
subshells inside the core region become larger than in the lower channels such 
as $3d$ transition metals and $4f$ elements. 
The DMC calculation difficulties come in form of biased expectations that stem from either too inaccurate trial functions or too crude approximation for the propagator whenever  local energy fluctuations become too large.

We study these issues on BeH$_2$ and C$_2$H$_2$, molecules that are simple and provide perhaps
the simplest illustrations how such behavior can be addressed.
Interestingly, we find that somewhat mundane issue 
of basis set accuracy could emerge as an important factor if the basis is not accurate enough.
Essentially, using a sufficiently accurate basis actually provides a straightforward way to address the difficulty.
This is true also for cases where the evaluation of nonlocal terms is carried out with trial functions that do not include the Jastrow factor.
Another
way to approach this 
for Slater-Jastrow 
wave functions is to employ T-moves that 
stabilize the calculations even with imperfect
basis and single-reference trial functions.
This can be used across all types of systems beyond molecules, such as solids with periodical boundary conditions.
We also show that the improvement of trial functions provides another route to address 
these difficulties. These calculations are all checked with CCSD(T)
and FCI basis set extrapolated 
calculations and the results show a high degree of consistency.

We can think about other cases where similar behavior can occur. 
For example,
near degeneracy effects 
might be encountered in calculations of ionized
or metallic systems in 3D.
Furthermore, this could be relevant also for calculations of semicore or core state excitations due to the different energy scales involved and 
the presence of a charged hole in the core region. 
Such states could exhibit very significant energy fluctuations, resulting in very low energy pockets. 
That could make stability of the runs and reliable expectation energies rather laborious. 
The presented work points out several options
on how to approach such complications.

The importance of the presented study  
is beyond just a necessary technical improvements.
The issue of nonlocal operators in electronic structure is difficult to avoid in general, mainly because of two reasons.
First, for statistical approaches 
the largest energy scale involved dominates the fluctuations.
Therefore, effective Hamiltonians
that eliminate such unimportant scales are crucial, but they typically involve nonlocal effective operators. 
Second, for heavy elements the nonlocal terms occur due to the fundamental nature of the problem as it is clear, for example, from the importance of spin-orbit interactions.  
From this perspective, methods that enable us to treat such Hamiltonians efficiently are crucial as they provide robust and broadly applicable approaches for solving electronic structure problems.

\section{Data Availibility}
Obtained results and data are presented in the Supplementary Material and in the main article.
Supplementary Material provides molecular geometries, molecular total energies for each basis set, as well as atomic total and kinetic energies for each basis set. 
Input and output files generated in this work are published in Materials Data Facility \cite{blaiszikMaterialsDataFacility2016, blaiszikDataEcosystemSupport2019} and can be found at Ref. \cite{mdf_data}.

\section{Declaration of Competing Interest}
The authors declare that they have no known competing financial interests or personal relationships that could have appeared to influence the work reported in this paper.

\section{Acknowledgements}
We gratefully acknowledge support by the U.S. Department of Energy, Office of Science, Basic Energy Sciences, Materials Sciences and Engineering Division, Theoretical Condensed Matter Physics under the award de-sc0012314.

\bibliographystyle{elsarticle-num} 
\bibliography{main}

\begin{thebibliography}{10}
\expandafter\ifx\csname url\endcsname\relax
  \def\url#1{\texttt{#1}}\fi
\expandafter\ifx\csname urlprefix\endcsname\relax\def\urlprefix{URL }\fi
\expandafter\ifx\csname href\endcsname\relax
  \def\href#1#2{#2} \def\path#1{#1}\fi

\bibitem{foulkes_quantum_2001}
W.~M.~C. Foulkes, L.~Mitas, R.~J. Needs, G.~Rajagopal, Quantum {Monte} {Carlo}
  simulations of solids, Rev. Mod. Phys. 73~(1) (2001) 33--83.
\newblock \href {https://doi.org/10.1103/RevModPhys.73.33}
  {\path{doi:10.1103/RevModPhys.73.33}}.

\bibitem{huntQuantumMonteCarlo2018}
R.~J. Hunt, M.~Szyniszewski, G.~I. Prayogo, R.~Maezono, N.~D. Drummond, Quantum
  {{Monte Carlo}} calculations of energy gaps from first principles, Physical
  Review B 98~(7) (2018) 075122.
\newblock \href {https://doi.org/10.1103/PhysRevB.98.075122}
  {\path{doi:10.1103/PhysRevB.98.075122}}.

\bibitem{kolorencApplicationsQuantumMonte2011}
J.~Koloren{\v c}, L.~Mitas, Applications of quantum {{Monte Carlo}} methods in
  condensed systems, Reports on Progress in Physics 74~(2) (2011) 026502.
\newblock \href {https://doi.org/10.1088/0034-4885/74/2/026502}
  {\path{doi:10.1088/0034-4885/74/2/026502}}.

\bibitem{al-hamdaniInteractionsLargeMolecules2021}
Y.~S. {Al-Hamdani}, P.~R. Nagy, A.~Zen, D.~Barton, M.~K{\'a}llay, J.~G.
  Brandenburg, A.~Tkatchenko, Interactions between large molecules pose a
  puzzle for reference quantum mechanical methods, Nature Communications 12~(1)
  (2021) 3927.
\newblock \href {https://doi.org/10.1038/s41467-021-24119-3}
  {\path{doi:10.1038/s41467-021-24119-3}}.

\bibitem{al-hamdaniWaterBNDoped2014}
Y.~S. {Al-Hamdani}, D.~Alf{\`e}, O.~A. {von Lilienfeld}, A.~Michaelides, Water
  on {{BN}} doped benzene: {{A}} hard test for exchange-correlation functionals
  and the impact of exact exchange on weak binding, The Journal of Chemical
  Physics 141~(18) (2014) 18C530.
\newblock \href {https://doi.org/10.1063/1.4898356}
  {\path{doi:10.1063/1.4898356}}.

\bibitem{dubeckyQuantumMonteCarlo2013}
M.~Dubeck{\'y}, P.~Jure{\v c}ka, R.~Derian, P.~Hobza, M.~Otyepka, L.~Mitas,
  Quantum {{Monte Carlo Methods Describe Noncovalent Interactions}} with
  {{Subchemical Accuracy}}, Journal of Chemical Theory and Computation 9~(10)
  (2013) 4287--4292.
\newblock \href {https://doi.org/10.1021/ct4006739}
  {\path{doi:10.1021/ct4006739}}.

\bibitem{wang_performance_2019}
T.~Wang, X.~Zhou, F.~Wang, Performance of the {Diffusion} {Quantum} {Monte}
  {Carlo} {Method} with a {Single}-{Slater}-{Jastrow} {Trial} {Wavefunction}
  {Using} {Natural} {Orbitals} and {Density} {Functional} {Theory} {Orbitals}
  on {Atomization} {Energies} of the {Gaussian}-2 {Set}, J. Phys. Chem. A
  123~(17) (2019) 3809--3817, publisher: American Chemical Society.
\newblock \href {https://doi.org/10.1021/acs.jpca.9b01933}
  {\path{doi:10.1021/acs.jpca.9b01933}}.

\bibitem{zhengComputationCorrelatedMetalInsulator2015}
H.~Zheng, L.~K. Wagner, Computation of the {{Correlated Metal}}-{{Insulator
  Transition}} in {{Vanadium Dioxide}} from {{First Principles}}, Physical
  Review Letters 114~(17) (2015) 176401.
\newblock \href {https://doi.org/10.1103/PhysRevLett.114.176401}
  {\path{doi:10.1103/PhysRevLett.114.176401}}.

\bibitem{huang_bandgaps_2021}
X.~Huang, H.~Zhang, X.-L. Cheng, Bandgaps in free-standing monolayer {TiO2}: Ab
  initio diffusion quantum monte carlo study, International Journal of Quantum
  Chemistry 121~(12) (2021) e26643.
\newblock \href {https://doi.org/https://doi.org/10.1002/qua.26643}
  {\path{doi:https://doi.org/10.1002/qua.26643}}.

\bibitem{wines_first-principles_2020}
D.~Wines, K.~Saritas, C.~Ataca, A first-principles {Quantum} {Monte} {Carlo}
  study of two-dimensional ({2D}) {GaSe}, J. Chem. Phys. 153~(15) (2020)
  154704, publisher: American Institute of Physics.
\newblock \href {https://doi.org/10.1063/5.0023223}
  {\path{doi:10.1063/5.0023223}}.

\bibitem{shinOptimizedStructureElectronic2021}
H.~Shin, J.~T. Krogel, K.~Gasperich, P.~R.~C. Kent, A.~Benali, O.~Heinonen,
  Optimized structure and electronic band gap of monolayer {{GeSe}} from
  quantum {{Monte Carlo}} methods, Physical Review Materials 5~(2) (2021)
  024002.
\newblock \href {https://doi.org/10.1103/PhysRevMaterials.5.024002}
  {\path{doi:10.1103/PhysRevMaterials.5.024002}}.

\bibitem{liAtomicFermiGas2011}
X.~Li, J.~Koloren{\v c}, L.~Mitas, Atomic {{Fermi}} gas in the unitary limit by
  quantum {{Monte Carlo}} methods: {{Effects}} of the interaction range,
  Physical Review A 84~(2) (2011) 023615.
\newblock \href {https://doi.org/10.1103/PhysRevA.84.023615}
  {\path{doi:10.1103/PhysRevA.84.023615}}.

\bibitem{bennett_new_2017}
M.~C. Bennett, C.~A. Melton, A.~Annaberdiyev, G.~Wang, L.~Shulenburger,
  L.~Mitas, A new generation of effective core potentials for correlated
  calculations, J. Chem. Phys. 147~(22) (2017) 224106.
\newblock \href {https://doi.org/10.1063/1.4995643}
  {\path{doi:10.1063/1.4995643}}.

\bibitem{bennettNewGenerationEffective2018}
M.~C. Bennett, G.~Wang, A.~Annaberdiyev, C.~A. Melton, L.~Shulenburger,
  L.~Mitas, A new generation of effective core potentials from correlated
  calculations: 2nd row elements, The Journal of Chemical Physics 149~(10)
  (2018) 104108.
\newblock \href {https://doi.org/10.1063/1.5038135}
  {\path{doi:10.1063/1.5038135}}.

\bibitem{annaberdiyevNewGenerationEffective2018}
A.~Annaberdiyev, G.~Wang, C.~A. Melton, M.~C. Bennett, L.~Shulenburger,
  L.~Mitas, A new generation of effective core potentials from correlated
  calculations: 3d transition metal series, The Journal of Chemical Physics
  149~(13) (2018) 134108.
\newblock \href {https://doi.org/10.1063/1.5040472}
  {\path{doi:10.1063/1.5040472}}.

\bibitem{wang_new_2019}
G.~Wang, A.~Annaberdiyev, C.~A. Melton, M.~C. Bennett, L.~Shulenburger,
  L.~Mitas, A new generation of effective core potentials from correlated
  calculations: 4s and 4p main group elements and first row additions, J. Chem.
  Phys. 151~(14) (2019) 144110, publisher: American Institute of Physics.
\newblock \href {https://doi.org/10.1063/1.5121006}
  {\path{doi:10.1063/1.5121006}}.

\bibitem{annaberdiyevAccurateAtomicCorrelation2020}
A.~Annaberdiyev, C.~A. Melton, M.~C. Bennett, G.~Wang, L.~Mitas, Accurate
  {{Atomic Correlation}} and {{Total Energies}} for {{Correlation Consistent
  Effective Core Potentials}}, Journal of Chemical Theory and Computation
  16~(3) (2020) 1482--1502.
\newblock \href {https://doi.org/10.1021/acs.jctc.9b00962}
  {\path{doi:10.1021/acs.jctc.9b00962}}.

\bibitem{mitas_nonlocal_1991}
L.~Mitáš, E.~L. Shirley, D.~M. Ceperley, Nonlocal pseudopotentials and
  diffusion {Monte} {Carlo}, J. Chem. Phys. 95~(5) (1991) 3467--3475,
  publisher: American Institute of Physics.
\newblock \href {https://doi.org/10.1063/1.460849}
  {\path{doi:10.1063/1.460849}}.

\bibitem{casula_beyond_2006}
M.~Casula, Beyond the locality approximation in the standard diffusion {Monte}
  {Carlo} method, Phys. Rev. B 74~(16) (2006) 161102, publisher: American
  Physical Society.
\newblock \href {https://doi.org/10.1103/PhysRevB.74.161102}
  {\path{doi:10.1103/PhysRevB.74.161102}}.

\bibitem{casula_size-consistent_2010}
M.~Casula, S.~Moroni, S.~Sorella, C.~Filippi, Size-consistent variational
  approaches to nonlocal pseudopotentials: {Standard} and lattice regularized
  diffusion {Monte} {Carlo} methods revisited, J. Chem. Phys. 132~(15) (2010)
  154113, publisher: American Institute of Physics.
\newblock \href {https://doi.org/10.1063/1.3380831}
  {\path{doi:10.1063/1.3380831}}.

\bibitem{caffarelUsingCIPSINodes2016}
M.~Caffarel, T.~Applencourt, E.~Giner, A.~Scemama, Using {{CIPSI Nodes}} in
  {{Diffusion Monte Carlo}}, in: Recent {{Progress}} in {{Quantum Monte
  Carlo}}, Vol. 1234 of {{ACS Symposium Series}}, {American Chemical Society},
  2016, Ch.~2, pp. 15--46.
\newblock \href {https://doi.org/10.1021/bk-2016-1234.ch002}
  {\path{doi:10.1021/bk-2016-1234.ch002}}.

\bibitem{anderson_nonlocal_2021}
T.~A. Anderson, C.~J. Umrigar, Nonlocal pseudopotentials and time-step errors
  in diffusion {Monte} {Carlo}, J. Chem. Phys. 154~(21) (2021) 214110,
  publisher: American Institute of Physics.
\newblock \href {https://doi.org/10.1063/5.0052838}
  {\path{doi:10.1063/5.0052838}}.

\bibitem{umrigar_diffusion_1993}
C.~J. Umrigar, M.~P. Nightingale, K.~J. Runge, A diffusion {Monte} {Carlo}
  algorithm with very small time‐step errors, J. Chem. Phys. 99~(4) (1993)
  2865--2890, publisher: American Institute of Physics.
\newblock \href {https://doi.org/10.1063/1.465195}
  {\path{doi:10.1063/1.465195}}.

\bibitem{dzubak}
A.~Dzubak, J.~Krogel, F.~Reboredo, Quantitative estimation of localization
  errors of 3d transition metal pseudopotentials in diffusion {M}onte {C}arlo,
  J. Chem. Phys. 147 (2017) 024102, publisher: American Institute of Physics.

\bibitem{krogelkent}
J.~Krogel, P.~Kent, Magnitude of pseudopotential localization errors in fixed
  node diffusion quantum monte carlo, J. Chem. Phys. 146 (2017) 244101,
  publisher: American Institute of Physics.

\bibitem{ginerUsingPerturbativelySelected2013}
E.~Giner, A.~Scemama, M.~Caffarel, Using perturbatively selected configuration
  interaction in quantum {{Monte Carlo}} calculations, Canadian Journal of
  Chemistry 91~(9) (2013) 879--885.
\newblock \href {https://doi.org/10.1139/cjc-2013-0017}
  {\path{doi:10.1139/cjc-2013-0017}}.

\bibitem{tenhaafProofUpperBound1995}
D.~F.~B. {ten Haaf}, H.~J.~M. {van Bemmel}, J.~M.~J. {van Leeuwen}, W.~{van
  Saarloos}, D.~M. Ceperley, Proof for an upper bound in fixed-node {{Monte
  Carlo}} for lattice fermions, Physical Review B 51~(19) (1995) 13039--13045.
\newblock \href {https://doi.org/10.1103/PhysRevB.51.13039}
  {\path{doi:10.1103/PhysRevB.51.13039}}.

\bibitem{kim_qmcpack_2018}
J.~Kim, A.~D. Baczewski, T.~D. Beaudet, A.~Benali, M.~C. Bennett, M.~A.
  Berrill, N.~S. Blunt, E.~J.~L. Borda, M.~Casula, D.~M. Ceperley, S.~Chiesa,
  B.~K. Clark, R.~C. Clay, K.~T. Delaney, M.~Dewing, K.~P. Esler, H.~Hao,
  O.~Heinonen, P.~R.~C. Kent, J.~T. Krogel, I.~Kylänpää, Y.~W. Li, M.~G.
  Lopez, Y.~Luo, F.~D. Malone, R.~M. Martin, A.~Mathuriya, J.~McMinis, C.~A.
  Melton, L.~Mitas, M.~A. Morales, E.~Neuscamman, W.~D. Parker, S.~D.~P.
  Flores, N.~A. Romero, B.~M. Rubenstein, J.~A.~R. Shea, H.~Shin,
  L.~Shulenburger, A.~F. Tillack, J.~P. Townsend, N.~M. Tubman, B.~V.~D. Goetz,
  J.~E. Vincent, D.~C. Yang, Y.~Yang, S.~Zhang, L.~Zhao, {QMCPACK}: an open
  sourceab initioquantum {Monte} {Carlo} package for the electronic structure
  of atoms, molecules and solids, J. Phys.: Condens. Matter 30~(19) (2018)
  195901, publisher: IOP Publishing.
\newblock \href {https://doi.org/10.1088/1361-648X/aab9c3}
  {\path{doi:10.1088/1361-648X/aab9c3}}.

\bibitem{kent_qmcpack_2020}
P.~R.~C. Kent, A.~Annaberdiyev, A.~Benali, M.~C. Bennett, E.~J. Landinez~Borda,
  P.~Doak, H.~Hao, K.~D. Jordan, J.~T. Krogel, I.~Kylänpää, J.~Lee, Y.~Luo,
  F.~D. Malone, C.~A. Melton, L.~Mitas, M.~A. Morales, E.~Neuscamman, F.~A.
  Reboredo, B.~Rubenstein, K.~Saritas, S.~Upadhyay, G.~Wang, S.~Zhang, L.~Zhao,
  {QMCPACK}: {Advances} in the development, efficiency, and application of
  auxiliary field and real-space variational and diffusion quantum {Monte}
  {Carlo}, J. Chem. Phys. 152~(17) (2020) 174105, publisher: American Institute
  of Physics.
\newblock \href {https://doi.org/10.1063/5.0004860}
  {\path{doi:10.1063/5.0004860}}.

\bibitem{krogel_nexus_2016}
J.~T. Krogel, Nexus: {A} modular workflow management system for quantum
  simulation codes, Computer Physics Communications 198 (2016) 154--168.
\newblock \href {https://doi.org/10.1016/j.cpc.2015.08.012}
  {\path{doi:10.1016/j.cpc.2015.08.012}}.

\bibitem{wagner_qwalk_2009}
L.~K. Wagner, M.~Bajdich, L.~Mitas, {QWalk}: {A} quantum {Monte} {Carlo}
  program for electronic structure, Journal of Computational Physics 228~(9)
  (2009) 3390--3404.
\newblock \href {https://doi.org/10.1016/j.jcp.2009.01.017}
  {\path{doi:10.1016/j.jcp.2009.01.017}}.

\bibitem{garnironQuantumPackageOpenSource2019}
Y.~Garniron, T.~Applencourt, K.~Gasperich, A.~Benali, A.~Fert{\'e}, J.~Paquier,
  B.~Pradines, R.~Assaraf, P.~Reinhardt, J.~Toulouse, P.~Barbaresco, N.~Renon,
  G.~David, J.-P. Malrieu, M.~V{\'e}ril, M.~Caffarel, P.-F. Loos, E.~Giner,
  A.~Scemama, Quantum {{Package}} 2.0: {{An Open}}-{{Source
  Determinant}}-{{Driven Suite}} of {{Programs}}, Journal of Chemical Theory
  and Computation 15~(6) (2019) 3591--3609.
\newblock \href {https://doi.org/10.1021/acs.jctc.9b00176}
  {\path{doi:10.1021/acs.jctc.9b00176}}.

\bibitem{scemama_qmc}
A.~Scemama, M.~Caffarel, E.~Oseret, W.~Jalby, {QMC=Chem: A Quantum Monte Carlo
  Program for Large-Scale Simulations in Chemistry at the Petascale Level and
  beyond}, in: {High Performance Computing for Computational Science - VECPAR
  2012}, Lecture Notes in Computer Science, {Springer Berlin Heidelberg}, 2013,
  pp. 118--127.
\newblock \href {https://doi.org/10.1007/978-3-642-38718-0\_14}
  {\path{doi:10.1007/978-3-642-38718-0\_14}}.

\bibitem{schmidt_general_1993}
M.~W. Schmidt, K.~K. Baldridge, J.~A. Boatz, S.~T. Elbert, M.~S. Gordon, J.~H.
  Jensen, S.~Koseki, N.~Matsunaga, K.~A. Nguyen, S.~Su, T.~L. Windus,
  M.~Dupuis, J.~A. Montgomery, General atomic and molecular electronic
  structure system, Journal of Computational Chemistry 14~(11) (1993)
  1347--1363.
\newblock \href {https://doi.org/10.1002/jcc.540141112}
  {\path{doi:10.1002/jcc.540141112}}.

\bibitem{sharma_spectroscopic_2014}
S.~Sharma, T.~Yanai, G.~H. Booth, C.~J. Umrigar, G.~K.-L. Chan, Spectroscopic
  accuracy directly from quantum chemistry: {Application} to ground and excited
  states of beryllium dimer, J. Chem. Phys. 140~(10) (2014) 104112, publisher:
  American Institute of Physics.
\newblock \href {https://doi.org/10.1063/1.4867383}
  {\path{doi:10.1063/1.4867383}}.

\bibitem{kalemos_nature_2016}
A.~Kalemos, The nature of the chemical bond in {Be$_2$}$^+$, {Be$_2$},
  {Be$_2$}$^-$, and {Be$_3$}, J. Chem. Phys. 145~(21) (2016) 214302, publisher:
  American Institute of Physics.
\newblock \href {https://doi.org/10.1063/1.4967819}
  {\path{doi:10.1063/1.4967819}}.

\bibitem{hachmannNodesHartreeFock2004}
J.~Hachmann, P.~T.~A. Galek, T.~Yanai, G.~K.-L. Chan, N.~C. Handy, The nodes of
  {{Hartree}}\textendash{{Fock}} wavefunctions and their orbitals, Chemical
  Physics Letters 392~(1) (2004) 55--61.
\newblock \href {https://doi.org/10.1016/j.cplett.2004.04.070}
  {\path{doi:10.1016/j.cplett.2004.04.070}}.

\bibitem{bandeRydbergStatesQuantum2006}
A.~Bande, A.~L{\"u}chow, F.~Della~Sala, A.~G{\"o}rling, Rydberg states with
  quantum {{Monte Carlo}}, The Journal of Chemical Physics 124~(11) (2006)
  114114.
\newblock \href {https://doi.org/10.1063/1.2180773}
  {\path{doi:10.1063/1.2180773}}.

\bibitem{raschCommunicationFixednodeErrors2014}
K.~M. Rasch, S.~Hu, L.~Mitas, Communication: {{Fixed}}-node errors in quantum
  {{Monte Carlo}}: {{Interplay}} of electron density and node nonlinearities,
  The Journal of Chemical Physics 140~(4) (2014) 041102.
\newblock \href {https://doi.org/10.1063/1.4862496}
  {\path{doi:10.1063/1.4862496}}.

\bibitem{burkatzkiEnergyconsistentPseudopotentialsQuantum2007}
M.~Burkatzki, C.~Filippi, M.~Dolg, Energy-consistent pseudopotentials for
  quantum {{Monte Carlo}} calculations, The Journal of Chemical Physics
  126~(23) (2007) 234105.
\newblock \href {https://doi.org/10.1063/1.2741534}
  {\path{doi:10.1063/1.2741534}}.

\bibitem{reynoldsFixedNodeQuantum1982a}
P.~J. Reynolds, D.~M. Ceperley, B.~J. Alder, W.~A. Lester, Fixed-node quantum
  {{Monte Carlo}} for molecules, The Journal of Chemical Physics 77~(11) (1982)
  5593--5603.
\newblock \href {https://doi.org/10.1063/1.443766}
  {\path{doi:10.1063/1.443766}}.

\bibitem{mitasWeightedNodalDomain2021}
L.~Mitas, A.~Annaberdiyev, Weighted nodal domain averages of eigenstates for
  quantum {{Monte Carlo}} and beyond, arXiv:2109.01734 [physics] (Sep. 2021).
\newblock \href {http://arxiv.org/abs/2109.01734} {\path{arXiv:2109.01734}}.

\bibitem{dubeckyNoncovalentInteractionsQuantum2016}
M.~Dubeck{\'y}, L.~Mitas, P.~Jure{\v c}ka, Noncovalent {{Interactions}} by
  {{Quantum Monte Carlo}}, Chemical Reviews 116~(9) (2016) 5188--5215.
\newblock \href {https://doi.org/10.1021/acs.chemrev.5b00577}
  {\path{doi:10.1021/acs.chemrev.5b00577}}.

\bibitem{pt_au}
L. {M}itas, to be published.

\bibitem{blaiszikMaterialsDataFacility2016}
B.~Blaiszik, K.~Chard, J.~Pruyne, R.~Ananthakrishnan, S.~Tuecke, I.~Foster, The
  {{Materials Data Facility}}: {{Data Services}} to {{Advance Materials Science
  Research}}, JOM 68~(8) (2016) 2045--2052.
\newblock \href {https://doi.org/10.1007/s11837-016-2001-3}
  {\path{doi:10.1007/s11837-016-2001-3}}.

\bibitem{blaiszikDataEcosystemSupport2019}
B.~Blaiszik, L.~Ward, M.~Schwarting, J.~Gaff, R.~Chard, D.~Pike, K.~Chard,
  I.~Foster, A data ecosystem to support machine learning in materials science,
  MRS Communications 9~(4) (2019) 1125--1133.
\newblock \href {https://doi.org/10.1557/mrc.2019.118}
  {\path{doi:10.1557/mrc.2019.118}}.

\bibitem{mdf_data}
H.~Zhou, A.~Scemama, G.~Wang, A.~Annaberdiyev, B.~Kincaid, M.~Caffarel,
  L.~Mitas, \href{https://doi.org/10.18126/0hei-jtub}{Dataset for ``{A quantum
  Monte Carlo study of systems with effective core potentials and node
  nonlinearities}"}, Materials Data Facility (2021).
\newblock \href {https://doi.org/10.18126/0HEI-JTUB}
  {\path{doi:10.18126/0HEI-JTUB}}.
\newline\urlprefix\url{https://doi.org/10.18126/0hei-jtub}

\end{thebibliography}






\vspace{3cm}
\appendix
\section{Supplementary Material}

Molecular geometries used in this work are provided in \tref{tab:geom}.
Corresponding molecular CCSD(T) and estimated FCI total energies are given in Tables \ref{tab:c2h2}, \ref{tab:beh2}, \ref{tab:beh2-reg}.

In addition, we are tabulating the total energies for Li and Be atoms with ccECP-reg potentials using increasingly accurate methods (HF, CISD, CCSD(T), and FCI) in \tref{reg}.
Corresponding atomic kinetic energies using HF, CISD, and FCI are given in \tref{reg,kin}.

\begin{table*}[!htbp]
\centering
\caption{Geometries of beryllium dihydride and acetylene molecules used in this work. Units are in Angstrom.}
\label{tab:geom}
\begin{tabular}{l|llll}
\hline
\hline
{} &         C1 &         C2 &         H1 &         H2  \\
\hline
C$_2$H$_2$ &  0.6013 &  -0.6013 &  1.6644 &  -1.6644 \\ \hline\hline
{} &       &     Be     &         H1&         H2  \\ \hline
BeH$_2$  &     &    0.0000     & 1.3264  & -1.3264 \\
\hline
\hline
\end{tabular}
\end{table*}

\begin{table*}[!htbp]
\centering
\caption{
Estimation of the exact energy [Ha] for C$_2$H$_2$ molecule with ccECP.
Values of post-HF methods correspond to correlation energies unless indicated otherwise.
CBS energy is extrapolated with $n\in \{T,Q,5,6\}$ using aug-cc-pCV$n$Z basis sets.
}
\label{tab:c2h2}
\begin{tabular}{l|llllll}
\hline
\hline
{} &         DZ &         TZ &         QZ &         5Z &         6Z &          CBS \\
\hline
UCCSD(T) &  -0.286860 &  -0.342848 &  -0.356438 &  -0.360132 &  -0.362579 &   -0.3647(11) \\
RHF      & -12.102941 & -12.114941 & -12.115970 & -12.115985 & -12.116165 &  -12.1162(1)  \\
Total    &            &            &            &            &            &  -12.4808(11) \\
\hline
CC bind  &            &            &            &            &            &  -0.646664 \\
Estim. FCI & -12.3916(1) & -12.4599(3) & -12.4745(3) &            &            &  -12.48174 \\

\hline
\hline
\end{tabular}
\end{table*}

\begin{table*}[!htbp]
\centering
\caption{
Estimation of the exact energy [Ha] for BeH$_2$ molecule with He-core ccECP.
Values of post-HF methods correspond to correlation energies unless indicated otherwise.
CBS energy is extrapolated with $n\in \{T,Q,5,6\}$ using aug-cc-pV$n$Z basis sets.
}
\label{tab:beh2}
\begin{tabular}{l|llllll}
\hline
\hline
{}           &        DZ &        TZ &        QZ &        5Z &        6Z &            CBS \\
\hline
UCCSD(T)     & -0.072213 & -0.078312 & -0.080059 & -0.080604 & -0.080815 &  -0.081087(11) \\
RHF          & -2.163346 & -2.164527 & -2.165039 & -2.165099 & -2.165109 &  -2.165112(11) \\
FCI-UCCSD(T) & -0.000195 & & & & & \\
Total        &           &           &           &           &           &  -2.246199(15) \\
\hline
CC bind      &           &           &           &           &           & -0.23596074  \\
Estim. FCI   & -2.235754 & -2.243034 & -2.245279 & -2.245660 & -2.245982 & -2.246394  \\

\hline
\hline
\end{tabular}
\end{table*}

\begin{table*}[!htbp]
\centering
\caption{
Estimation of the exact energy [Ha] for BeH$_2$ molecule with ccECP-reg.
Values of post-HF methods correspond to correlation energies unless indicated otherwise.
CBS energy is extrapolated with $n\in \{T,Q,5\}$ using aug-cc-pCV$n$Z basis sets.
}
\label{tab:beh2-reg}
\begin{tabular}{l|llllll}
\hline
\hline
{} &         DZ &         TZ &         QZ &         5Z &        CBS \\
\hline
UCCSD(T) &  -0.106433 &  -0.121982 &  -0.126953 &  -0.128257 &  -0.129413 \\
RHF      & -15.772368 & -15.773465 & -15.774003 & -15.774036 & -15.774054 \\
Total    &            &            &             & & -15.903467 \\
\hline
CC bind  &            &            &          &            &  -0.2359234 \\
Estim. FCI & -15.844900 & -15.867966 & -15.878771 & -15.886439 & -15.903662  \\

\hline
\hline
\end{tabular}
\end{table*}

\begin{table*}[htbp!]
\setlength{\tabcolsep}{4pt} 
\centering
\small
\caption{
Atomic correlation and total energies [Ha] for ccECP(reg) pseudo-atoms. 
aug-cc-pCV$n$Z basis set. CBS value is from TZ-5Z data extrapolation.
}

\label{reg}
\begin{tabular}{l|l|rrrr|r}
\hline\hline
Atom & Method &          DZ &          TZ &          QZ &          5Z &         CBS \\
\hline
\multirow{6}{*}{\bf{Li}}
& CISD     & -0.0332009 & -0.0414622 & -0.0435773 & -0.0442375 &   -0.04484425 \\
& RCCSD(T) & -0.0332048 & -0.0414857 & -0.0436132 & -0.0442769 &   -0.04488658 \\
& UCCSD(T) & -0.0332048 & -0.0414860 & -0.0436135 & -0.0442773 &   -0.04488693 \\
& FCI      & -0.0332055 & -0.0414895 & -0.0436184 & -0.0442820 &   -0.04489113 \\
& ROHF     & -7.4332845 & -7.4332845 & -7.4332848 & -7.4332853 &   -7.43328591 \\
\cline{2-7}
& Total    &            &            &            &            &   -7.47817705 \\
\hline 

\multirow{6}{*}{\bf{Be}}
& CISD     &  -0.0760900 &  -0.0854581 &  -0.0882217 &  -0.0889584 &    -0.0894789 \\ 
& RCCSD(T) &  -0.0790710 &  -0.0891582 &  -0.0921943 &  -0.0929903 &    -0.0935334 \\ 
& UCCSD(T) &  -0.0790710 &  -0.0891582 &  -0.0921943 &  -0.0929904 &    -0.0935334 \\ 
& CCSDT(Q) &  -0.0790829 &  -0.0891843 &  -0.0922261 &  -0.0930201 &    -0.0935565 \\
& FCI      &  -0.0790833 &  -0.0891849 &  -0.0922266 &  -0.0930206 &    -0.0935569 \\
& RHF      & -14.5739828 & -14.5739830 & -14.5739830 & -14.5739845 &   -14.5739866 \\ 
& Total    &             &             &             &             &   -14.6675436 \\ 

\hline\hline
\end{tabular}
\end{table*}

\begin{table*}[htbp!]
\setlength{\tabcolsep}{4pt} 
\centering
\small
\caption{
ccECP(reg) kinetic energies [Ha]. aug-cc-pCV$n$Z basis set.
}
\label{reg,kin}
\begin{tabular}{l|l|rrrr|r}
\hline\hline
Atom & Method &          DZ &          TZ &          QZ &          5Z &             CBS \\
\hline

\multirow{3}{*}{\bf{Li}}
& ROHF & 7.19117556 & 7.19123544 & 7.19120893 & 7.19125638 &            \\
& CISD & 7.21584654 & 7.22815685 & 7.23145895 & 7.23240444 &            \\
& FCI  & 7.21606921 & 7.22899628 & 7.23252884 & 7.23354102 &  7.2346(5) \\
\hline

\multirow{3}{*}{\bf{Be}}
& RHF  & 14.05600333 & 14.05584399 & 14.05606223 & 14.05614560 &             \\
& CISD & 14.12678062 & 14.13517666 & 14.13987238 & 14.14109025 &             \\
& FCI  & 14.13040123 & 14.14067897 & 14.14628457 & 14.14769193 &  14.1491(7) \\

\hline\hline
\end{tabular}
\end{table*}

\end{document}